\documentstyle[prl,aps,preprint]{revtex}

\begin{document}
\title{Shear-free rotating inflation}

\author{Yuri N.\ Obukhov\footnote{On leave from: Department
of Theoretical Physics, Moscow State University, 117234 Moscow,
Russia}, Thoralf Chrobok, and Mike Scherfner\footnote{FB Mathematik,
Technische Universit\"at Berlin, Str. d. 17. Juni 136, D-10623
Berlin}}
\address
{Institute of Theoretical Physics, Technische Universit\"at Berlin,
Hardenbergstr. 36, D-10623 Berlin, Germany}

\maketitle

\begin{abstract}
We demonstrate the existence of shear-free cosmological models
with rotation and expansion which support the inflationary
scenarios. The corresponding metrics belong to the family of
spatially homogeneous models with the geometry of the closed
universe (Bianchi type IX). We show that the global vorticity does
not prevent the inflation and even can accelerate it.
\end{abstract}

\pacs{PACS no.: 98.80.Cq; 04.20.Jb; 04.20.-q; 98.80.Hw} 

\section{Introduction}

Rotation is a universal physical phenomenon. All known objects from
the fundamental particles to planets, stars, and galaxies are rotating.
We then naturally come to the question whether the largest physical
system -- the universe -- has such a property. This problem comprises
several aspects: If our world does not rotate, then why and how this
happens? Since the rotating models cannot be excluded from consideration
a priori, it is necessary to reveal a physical mechanism which prevents 
the universal rotation. On the other hand, if the world can and does 
rotate, then what are the corresponding observational manifestations 
of the cosmic rotation? Technically, this reduces to the study of the 
geometry of a rotating cosmological model and to the analysis of the
motion of particles and light in such a spacetime manifold. And the 
ultimate question is of course about the dynamical realization of 
the rotating models, i.e. the description of the realistic matter
sources and the derivation of the solutions of the gravitational field
equations. 

Since the early works of Lanczos\cite{lanc}, Gamov \cite{gamov}, and 
G\"odel \cite{goe}, the cosmological models with rotation have been
studied in a great number of publications (see the overview in \cite{rotrev} 
and the exhaustive list of references therein). Quite strong upper limits 
for the cosmic vorticity were obtained from the analysis of the observed 
properties of the microwave background radiation \cite{col}. However, all 
these works deal with the models in which shear and vorticity are 
inseparable (in the sense that zero shear automatically implies zero 
vorticity). Correspondingly, the limits \cite{col} are actually placed 
not on the vorticity, but rather on the shear induced by it within the 
specific geometrical models. One thus needs a separate analysis of the
cosmological models with trivial shear but nonzero rotation and expansion. 

Earlier \cite{rotrev} we have studied the wide class of spatially 
homogeneous models described by the metric
\begin{equation}
ds^2 = dt^2 - 2\,R\,n_{a}dx^{a}dt -
R^2\,\gamma_{ab}\,dx^{a}dx^{b}.\label{met0}
\end{equation}
Here the indices $a,b,c = 1,2,3$ label the spatial coordinates, $R =
R(t)$ is the scale factor, and
\begin{equation}
n_{a}=\nu_A\,e_a^{A},\qquad
\gamma_{ab}=\beta_{AB}\,e_a^{A}e_b^{B},\label{ng}
\end{equation}
with the constant coefficients $\nu_{A},\beta_{AB}$ ($A,B = 1,2,3$).
The 1-forms $e^{A}= e_{a}^{A}(x)\,dx^a$ are invariant with respect to 
the action of a three-parameter group of motion which is admitted by 
the space-time (\ref{met0}). The action of this group is simply-transitive 
on the spatial ($t=const$) hypersurfaces. There exist 9 types (Bianchi 
types) of such manifolds, distinguished by the Killing vectors $\xi_{A}$ 
and their commutators $[\xi_{A},\xi_{B}]=f^{C}{}_{AB}\,\xi_{C}$.

Models (\ref{met0}) are shear-free but the vorticity and expansion are 
nontrivial, in general. The {\it kinematic} analysis \cite{rotrev} of 
the models (\ref{met0}) reveals their several attractive properties: the 
complete causality (no timelike closed curves), the absence of parallax 
effects, and the isotropy of the microwave background radiation. As a
result, these shear-free models satisfy all the known observational 
criteria for the cosmic rotation. In particular, it is worthwhile to note 
that the vorticity bounds \cite{col} are not applicable to the class of 
metrics (\ref{met0}). The satisfactory observational properties suggest 
that the shear-free homogeneous models can be considered as the viable 
candidates for the description of the cosmic rotation. 

The aim of the present paper is to address the {\it dynamic} aspect
of the theory: namely, to study the realization of the models (\ref{met0}) 
as exact solutions of the gravitational field equations. This represents 
a nontrivial problem, in general, as it is notoriously difficult to 
combine the expansion with vorticity in a realistic cosmological model. 
In technical terms, the most important thing needed is to determine the 
physically reasonable matter content of such cosmologies.

In this paper we continue the study of the Bianchi type IX models
belonging to the class (\ref{met0}). The Bianchi IX type is distinguished
among the other spatially homogeneous models by the fact that its geometry
describes the spatially {\it closed} world. Many very interesting questions
related to the Mach principle arise in this connection. In particular, 
it is a matter of principal importance to know whether Einstein's field 
equations admit the truly anti-Machian solutions or not. A first example
of such a solution was given by the stationary model of Ozsv\'ath and 
Sch\"ucking \cite{osch}. However, later its anti-Machian nature was questioned 
by King \cite{king} who developed the idea that the total angular velocity
of the closed world is ultimately zero because the cosmic vorticity is 
compensated by the rotating gravitational waves. Recently, we have 
demonstrated the existence of another stationary rotating closed Bianchi
IX world in which the cosmic vorticity is balanced by the spin of the 
cosmological matter \cite{nine}. 

The above mentioned results refer to the stationary models which are
clearly of the academic interest only because of the absence of expansion.
Here we give two explicit examples of the more physically realistic
{\it non}stationary closed Bianchi IX worlds with nontrivial rotation 
and expansion. After the description of the spacetime geometry in 
Sec.~\ref{geometry}, we present the rotating version of the de Sitter 
solution in Sec.~\ref{desitter}. Further, in Sec.~\ref{inflation} we 
demonstrate that the shear-free models with rotation and expansion arise 
in the standard inflationary scheme.

\section{Closed world geometry}\label{geometry}

Closed spatially homogeneous Bianchi type IX worlds are
constructed with the help of the triad of invariant 1-forms
$e^{A}$ which satisfy the structure equations
\begin{equation}
de^A = f^A{}_{BC}\,e^B\wedge e^C,\quad {\rm with}\quad f^1{}_{23} =
f^2{}_{31} = f^3{}_{12} = 1.
\end{equation}
Denoting the spatial coordinates $x=x^1, y=x^2, z=x^3$, one can
choose them in the following explicit realization:
\begin{eqnarray}
e^1 &=& \cos y\,\cos z\,dx - \sin z\,dy,\nonumber\\ e^2 &=& \cos
y\,\sin z\,dx + \cos z\,dy,\nonumber\\ e^3 &=& -\,\sin y\,dx +
dz.\label{eA}
\end{eqnarray}
We assume the diagonal $\beta_{AB}$ and can write the ansatz for
the line element (\ref{met0}) as
\begin{equation}
ds^2 = g_{\alpha\beta}\,\vartheta^\alpha\,\vartheta^\beta, \qquad
g_{\alpha\beta} = {\rm diag}(1, -1, -1, -1),\label{met1}
\end{equation}
where the orthonormal coframe 1-forms $\vartheta^\alpha$ read
\begin{equation}
\vartheta^{\widehat{0}} = dt - R\,\nu_A\,e^A,\
\vartheta^{\widehat{1}} = R\,k_1\,e^1,\ \vartheta^{\widehat{2}} =
R\,k_2\,e^2,\ \vartheta^{\widehat{3}} = R\,k_3\,e^3.\label{var}
\end{equation}
Here, $k_1, k_2, k_3$ are positive constant parameters. The Greek
indices $\alpha,\beta,\dots = 0,1,2,3$ hereafter label the objects
with respect to the orthonormal frame; the hats over indices denote
the separate frame components of these objects.

The kinematic properties of the spacetime geometry are described
by the vorticity $\omega_{\mu\nu} =
h^{\alpha}{}_{\mu}h^{\beta}{}_{\nu} \nabla_{[\alpha}u_{\beta]}$,
shear $\sigma_{\mu\nu}=h^{\alpha}{}_{\mu}
h^{\beta}{}_{\nu}\nabla_{(\alpha}u_{\beta)} - {1\over
3}\,h_{\mu\nu} \nabla_{\lambda}u^{\lambda}$, and the volume
expansion $\theta= \nabla_{\lambda}u^{\lambda}$. Here
$u=\partial_t$ is the comoving velocity (normalized by $u_\alpha u
^\alpha =1$) and $h_{\mu\nu} = g_{\mu\nu} - u_\mu u_\nu$ is the
standard projector on the rest 3-space. A direct calculation
yields:
\begin{eqnarray}
&& \sigma_{\mu\nu} = 0,\qquad a^{\widehat{1}} = {\frac
{\dot{R}\nu_1}{Rk_1}},\quad a^{\widehat{2}} = {\frac
{\dot{R}\nu_2}{Rk_2}},\quad a^{\widehat{3}} = {\frac
{\dot{R}\nu_3}{Rk_3}},\qquad \theta = 3{\frac {\dot{R}}{R}},\\ &&
\omega_{\widehat{2}\widehat{3}} = -\,{\frac {\nu_1}{2Rk_2k_3}},\qquad
\omega_{\widehat{3}\widehat{1}} = -\,{\frac {\nu_2}{2Rk_1k_3}},\qquad
\omega_{\widehat{1}\widehat{2}} = -\,{\frac
{\nu_3}{2Rk_1k_2}}.\label{kin}
\end{eqnarray}

\section{Rotating de Sitter world}\label{desitter}

First we study the case when matter is represented by just the
cosmological constant. An equivalent physical model is given by
the ideal fluid with the vacuum equation of state. The total
Lagrangian reads
\begin{equation}
L = -\,{\frac 1 {2\kappa}}\,(R + 2\Lambda),
\end{equation}
and the left-hand side of the Einstein field equations,
$R_{\alpha\beta} - {\frac 1 2}\,R\,g_{\alpha\beta} =
\Lambda\,g_{\alpha\beta}$, take the form (\ref{G00})-(\ref{G23})
given in the Appendix.

As a first step, we specialize to the case
\begin{equation}
\nu_1\neq 0, \qquad \nu_2 = \nu_3 = 0.\label{nu}
\end{equation}
Then there remains only the ``01'' nontrivial off-diagonal
equation which reduces to
\begin{equation}
-\,{\frac {\ddot{R}}{R}} + {\frac {\dot{R}^2}{R^2}} + {\frac
{k_1^2}{4R^2\,k_2^2k_3^2}} =0.\label{off1}
\end{equation}
The analysis of the four diagonal Einstein equations, see
(\ref{G00})-(\ref{G33}), shows that they are consistent under the
algebraic conditions
\begin{equation}
k_3 = k_2,\qquad {\rm and}\qquad k_2^2 = k_1^2 - \nu_1^2.\label{k23}
\end{equation}
Then the diagonal equations, using (\ref{off1}), reduce to the first
order equation
\begin{equation}
3\left({\frac {\dot{R}^2}{R^2}} + {\frac {k_1^2}
{4R^2\,k_2^4}}\right) = {\frac {k_1^2}{k_2^2}}\,\Lambda.
\end{equation}
This can be straightforwardly integrated, yielding the solution
\begin{equation}
R(t) = {\frac 1{2k_2}}\,\sqrt{\frac 3 \Lambda}\,\cosh\left( {\frac
{k_1}{k_2}}\,\sqrt{\frac \Lambda 3}\,t\right).\label{des}
\end{equation}
One can check that (\ref{off1}) is then identically fulfilled. The
metric (\ref{met1}), (\ref{var}) with the scale factor (\ref{des})
represents the rotating version of the de Sitter world. A slightly
different form of that solution was obtained in \cite{sacha}. Another
rotating generalization of the de Sitter model is described in
\cite{gron} which is also the shear-free Bianchi type IX, although it
does not belong to the class (\ref{met0}).

\section{Rotating inflationary models}\label{inflation}

Inspired by the above preliminary demonstration that rotation can
coexist with inflation, we now consider the general inflationary
model (see \cite{infrev,linde,guth,liddle}, for example) which is
described by the Lagrangian with the scalar field
\begin{equation}
L = -\,{\frac 1 {2\kappa}}\,R + {\frac 1 2}\,(\partial_\mu\phi)
(\partial^\mu\phi) - V(\phi).
\end{equation}
The Einstein equations now read $R_{\alpha\beta} - {\frac 1 2}\,R
\,g_{\alpha\beta} = \kappa\,T_{\alpha\beta}$, where the explicit
form of the energy-momentum components is given in the Appendix,
see (\ref{T00})-(\ref{Tab}).

Again specializing to the case (\ref{nu}), we find that all the
off-diagonal equations are trivially fulfilled except for the ``01''
component. The latter reads:
\begin{equation}
2\left(-\,{\frac {\ddot{R}}{R}} + {\frac {\dot{R}^2}{R^2}} + {\frac
{k_1^2}{4R^2\,k_2^2k_3^2}}\right) =\kappa\,\dot{\phi}^2.\label{off2}
\end{equation}
Substituting $\kappa\,\dot{\phi}^2$ from (\ref{off2}) into the four
diagonal Einstein equations [use (\ref{G00})-(\ref{G33}) and
(\ref{T00})-(\ref{T33})], we again discover the consistency condition
(\ref{k23}). As a result, the diagonal equations reduce to
\begin{equation}
{\frac {\ddot{R}}{R}} + 2\,{\frac {\dot{R}^2}{R^2}} + {\frac {k_1^2}
{2R^2\,k_2^4}} = {\frac {k_1^2}{k_2^2}}\,\kappa V.\label{diag}
\end{equation}

Beside the Einstein equations, we have the Klein-Gordon equation
for the scalar field, $D_\mu D^\mu\phi + V' = 0$ [where
$V':=dV(\phi)/d\phi$]. For the metric (\ref{met1})-(\ref{var}) it
reads:
\begin{equation}
\ddot{\phi} + 3\,{\frac {\dot{R}}{R}}\,\dot{\phi} + {\frac
{k_1^2}{k_2^2}}\,V' =0.\label{eq3}
\end{equation}

Only two of the three equations (\ref{off2})-(\ref{eq3}) are
independent. In order to see this, let us take, instead of
(\ref{off2}) and (\ref{diag}), their sum and difference. This yields
\begin{eqnarray}
{\frac {\dot{R}^2}{R^2}} + {\frac {k_1^2}{4R^2\,k_2^4}} = {\frac
\kappa 3}\left({\frac 1 2}\,\dot{\phi}^2 + {\frac
{k_1^2}{k_2^2}}\,V\right),\label{infl1}\\ {\frac {\ddot{R}}{R}} =
{\frac \kappa 3}\left(-\,\dot{\phi}^2 + {\frac
{k_1^2}{k_2^2}}\,V\right).\label{infl2}
\end{eqnarray}
We can take as the independent dynamical equations either
(\ref{infl1}) together with (\ref{eq3}), or (\ref{infl1}) together
with (\ref{infl2}). Then, correspondingly, the third equation will be
derived from the first two, provided $\dot{\phi}\neq 0$.

We thus have recovered the system of the usual inflationary model
in which the spatial curvature $K$ and the inflaton potential are
``corrected'' by the rotation parameters
\begin{equation}
K\longrightarrow {\frac {k_1^2}{4k_2^4}},\qquad V\longrightarrow
{\frac {k_1^2}{k_2^2}}\,V.\label{redef}
\end{equation}
The form of the exact or approximate solutions of the final system
depends on the inflaton potential $V(\phi)$, and we refer to the
relevant analysis of the standard inflationary system
\cite{infrev,linde,guth,liddle} (see also \cite{schunck} and
references therein) which are completely applicable to our
rotating world after we make the redefinitions (\ref{redef}).

\section{Discussion and conclusion}

The results of Sec.~\ref{desitter} represent a particular case of the
general inflationary model when $\dot{\phi}=0$ with $V$ playing the
role of the cosmological constant. However we found it more
instructive to consider that special case separately, in particular
because then it is possible to make a direct comparison with the
earlier results of \cite{gron}. With an account of the algebraic
conditions (\ref{nu}) and (\ref{k23}), we have constructed the exact
solution of the Einstein equations in the form of the line element
\begin{equation}
ds^2 = dt^2 - 2\nu_1\,R\,dt\,e^1 - k_2^2\,R^2\left[(e^1)^2 + (e^2)^2
+ (e^3)^2\right],\label{met2}
\end{equation}
where the scale factor $R$ is determined from (\ref{des}) or from 
the inflationary system (\ref{eq3})-(\ref{infl2}). This model is
shear-free, and the results obtained are thus contributing to the studies 
of the shear-free conjecture, see \cite{seno}, e.g. The volume expansion
is $\theta = 3\dot{R}/R$ and the vorticity is decreasing in the
expanding universe with the only nontrivial component
\begin{equation}
\omega_{\widehat{2}\widehat{3}} = -\,{\frac {\nu_1}{2Rk_2^2}}.
\end{equation}
During the de Sitter era (\ref{des}) the cosmic rotation rapidly decays.

Our results confirm and extend the conclusions of Gr{\o}n \cite{gron}, 
see also \cite{ell,gron3}, on that the cosmic rotation does not prevent 
the inflation, whereas the latter yields a quick decrease of vorticity. 
The preliminary and qualitative conclusions of \cite{ell,gron3} were 
derived on the basis of the conservation law of the angular momentum 
without analyzing Einstein's equations. The behavior of our exact 
solution provides now the direct evidence in support of these results.
Moreover, because of (\ref{redef}), we can see now that the cosmic 
vorticity in fact enhances the inflation: when the vorticity is large 
($\nu_1\rightarrow\infty$ for the fixed value of $k_2$) the coefficient 
$k_1/k_2>1$ makes the inflation rate much bigger than in the vorticity-free 
case ($k_1/k_2=1$ for $\nu_1=0$). 

Summarizing, we have demonstrated the existence of the realistic
cosmological model with rotation and expansion: The exact Bianchi IX 
solution (\ref{met2}) is determined by the standard inflationary system 
(\ref{infl1})-(\ref{infl2}). Here we do not specify the explicit form 
of the inflaton potential which represents a separate complicate subject 
in the modern cosmology. However for each given $V(\phi)$ the evolution of 
the scalar field and the cosmological scale factor can be straightforwardly
found. 

In our final remark, let us come back to the Mach principle. Since our model 
describes the {\it closed} world, its existence again raises the question 
whether the true anti-Mach cosmology is possible. The earlier discussion 
of the stationary models \cite{osch,king,nine} has revealed some mechanisms 
of compensation of the global vorticity by the gravitational wave or by the
local spin of matter. As far as we can see, such a compensation does not
exist for the new solution. This means that the shear-free rotating 
inflational Bianchi IX model describes the true (and far more realistic 
due to the nontrivial expansion) anti-Machian model. In this connection,
it would be also interesting to study the Bianchi type V rotating models 
which contain the open standard cosmology as a particular case.

\section{Acknowledgments}
This work was supported by the Deutsche Forschungsgemeinschaft with
the grant 436 RUS 17/70/01.

\section*{Appendix}

The left-hand side of the Einstein gravitational field equations is
described by the Einstein tensor $G_{\alpha\beta} = R_{\alpha\beta} -
{\frac 1 2}R\,g_{\alpha\beta}$. For the metric
(\ref{met1})-(\ref{var}),
 it reads:
\begin{eqnarray}
G_{\widehat{0}\widehat{0}} &=& -\left(2{\frac {\ddot{R}}{R}} + {\frac
{\dot{R}^2}{R^2}}\right)\left({\frac {\nu_1^2}{k_1^2}} + {\frac
{\nu_2^2}{k_2^2}} + {\frac {\nu_3^2}{k_3^2}}\right) + 3{\frac
{\dot{R}^2}{R^2}}\nonumber\\ &&+\,{\frac {-k_1^4 - k_2^4 - k_3^4 +
2(k_1^2k_2^2 + k_1^2k_3^2 + k_2^2k_3^2) + 3(k_1^2\nu_1^2 +
k_2^2\nu_2^2 + k_3^2\nu_3^2)} {4R^2\,(k_1k_2k_3)^2}},\label{G00}\\
G_{\widehat{1}\widehat{1}} &=& \left(2{\frac {\ddot{R}}{R}} + {\frac
{\dot{R}^2}{R^2}}\right)\left(-1 + {\frac {\nu_2^2}{k_2^2}} + {\frac
{\nu_3^2}{k_3^2}}\right) + 3{\frac {\dot{R}^2}{R^2}} \,{\frac
{\nu_1^2}{k_1^2}} \nonumber\\ &&+\,{\frac {3k_1^4 - k_2^4 - k_3^4 +
2(-k_1^2k_2^2 - k_1^2k_3^2 + k_2^2k_3^2) - k_1^2\nu_1^2 +
k_2^2\nu_2^2 + k_3^2\nu_3^2} {4R^2\,(k_1k_2k_3)^2}},\label{G11}\\
G_{\widehat{2}\widehat{2}} &=& \left(2{\frac {\ddot{R}}{R}} + {\frac
{\dot{R}^2}{R^2}}\right)\left(-1 + {\frac {\nu_1^2}{k_1^2}} + {\frac
{\nu_3^2}{k_3^2}}\right) + 3{\frac {\dot{R}^2}{R^2}} \,{\frac
{\nu_2^2}{k_2^2}} \nonumber\\ &&+\,{\frac {-k_1^4 + 3k_2^4 - k_3^4 +
2(-k_1^2k_2^2 + k_1^2k_3^2 - k_2^2k_3^2) + k_1^2\nu_1^2 -
k_2^2\nu_2^2 + k_3^2\nu_3^2} {4R^2\,(k_1k_2k_3)^2}},\label{G22}\\
G_{\widehat{3}\widehat{3}} &=& \left(2{\frac {\ddot{R}}{R}} + {\frac
{\dot{R}^2}{R^2}}\right)\left(-1 + {\frac {\nu_1^2}{k_1^2}} + {\frac
{\nu_2^2}{k_2^2}}\right) + 3{\frac {\dot{R}^2}{R^2}} \,{\frac
{\nu_3^2}{k_3^2}} \nonumber\\ &&+\,{\frac {-k_1^4 - k_2^4 + 3k_3^4 +
2(k_1^2k_2^2 - k_1^2k_3^2 - k_2^2k_3^2) + k_1^2\nu_1^2 + k_2^2\nu_2^2
- k_3^2\nu_3^2} {4R^2\,(k_1k_2k_3)^2}},\label{G33}\\
G_{\widehat{0}\widehat{1}} &=& 2\left(-\,{\frac {\ddot{R}}{R}} +
{\frac {\dot{R}^2}{R^2}}\right){\frac {\nu_1}{k_1}} + {\frac
{\dot{R}}{R^2}}\,{\frac {k_1\nu_2\nu_3(k_3^2 -
k_2^2)}{(k_1k_2k_3)^2}} + {\frac
{k_1\nu_1}{2R^2\,k_2^2k_3^2}},\label{G01}\\
G_{\widehat{0}\widehat{2}} &=& 2\left(-\,{\frac {\ddot{R}}{R}} +
{\frac {\dot{R}^2}{R^2}}\right){\frac {\nu_2}{k_2}} + {\frac
{\dot{R}}{R^2}}\,{\frac {k_2\nu_1\nu_3(k_1^2 -
k_3^2)}{(k_1k_2k_3)^2}} + {\frac
{k_2\nu_2}{2R^2\,k_1^2k_3^2}},\label{G02}\\
G_{\widehat{0}\widehat{3}} &=& 2\left(-\,{\frac {\ddot{R}}{R}} +
{\frac {\dot{R}^2}{R^2}}\right){\frac {\nu_3}{k_3}} + {\frac
{\dot{R}}{R^2}}\,{\frac {k_3\nu_1\nu_2(k_2^2 -
k_1^2)}{(k_1k_2k_3)^2}} + {\frac
{k_3\nu_3}{2R^2\,k_1^2k_2^2}},\label{G03}\\
G_{\widehat{1}\widehat{2}} &=& 2\left(-\,{\frac {\ddot{R}}{R}} +
{\frac {\dot{R}^2}{R^2}}\right){\frac {\nu_1\nu_2}{k_1k_2}} + {\frac
{\dot{R}}{R^2}}\,{\frac {k_1k_2\nu_3(k_1^2 - k_2^2)}{(k_1k_2k_3)^2}}
- {\frac {\nu_1\nu_2}{2R^2\,k_1k_2k_3^2}},\label{G12}\\
G_{\widehat{1}\widehat{3}} &=& 2\left(-\,{\frac {\ddot{R}}{R}} +
{\frac {\dot{R}^2}{R^2}}\right){\frac {\nu_1\nu_3}{k_1k_3}} + {\frac
{\dot{R}}{R^2}}\,{\frac {k_1k_3\nu_2(k_3^2 - k_1^2)}{(k_1k_2k_3)^2}}
- {\frac {\nu_1\nu_3}{2R^2\,k_1k_2^2k_3}},\label{G13}\\
G_{\widehat{2}\widehat{3}} &=& 2\left(-\,{\frac {\ddot{R}}{R}} +
{\frac {\dot{R}^2}{R^2}}\right){\frac {\nu_2\nu_3}{k_2k_3}} + {\frac
{\dot{R}}{R^2}}\,{\frac {k_2k_3\nu_1(k_2^2 - k_3^2)}{(k_1k_2k_3)^2}}
- {\frac {\nu_2\nu_3}{2R^2\,k_1^2k_2k_3}}.\label{G23}
\end{eqnarray}

The right-hand side of Einstein's equations is the energy-momentum
tensor of the scalar field $T_{\alpha\beta} =
(D_\alpha\phi)(D_\beta\phi) - {\frac 1
2}(D_\mu\phi)(D^\mu\phi)\,g_{\alpha\beta} + V\,g_{\alpha\beta}$.
For the metric (\ref{met1})-(\ref{var}), we find:
\begin{eqnarray}
T_{\widehat{0}\widehat{0}} &=& {\frac {\dot{\phi}^2}2}\left( 1 +
{\frac {\nu_1^2}{k_1^2}} + {\frac {\nu_2^2}{k_2^2}} + {\frac
{\nu_3^2}{k_3^2}}\right) + V,\label{T00}\\ T_{\widehat{1}\widehat{1}}
&=& {\frac {\dot{\phi}^2}2}\left( 1 + {\frac {\nu_1^2}{k_1^2}} -
{\frac {\nu_2^2}{k_2^2}} - {\frac {\nu_3^2}{k_3^2}}\right) -
V,\label{T11}\\ T_{\widehat{2}\widehat{2}} &=& {\frac
{\dot{\phi}^2}2}\left( 1 - {\frac {\nu_1^2}{k_1^2}} + {\frac
{\nu_2^2}{k_2^2}} - {\frac {\nu_3^2}{k_3^2}}\right) - V,\label{T22}\\
T_{\widehat{3}\widehat{3}} &=& {\frac {\dot{\phi}^2}2}\left( 1 -
{\frac {\nu_1^2}{k_1^2}} - {\frac {\nu_2^2}{k_2^2}} + {\frac
{\nu_3^2}{k_3^2}}\right) - V,\label{T33}\\ T_{\widehat{0}\widehat{1}}
&=& \dot{\phi}^2\,{\frac {\nu_1}{k_1}},\qquad
T_{\widehat{0}\widehat{2}}=\dot{\phi}^2\,{\frac {\nu_2}{k_2}},\qquad
T_{\widehat{0}\widehat{3}}=\dot{\phi}^2\,{\frac
{\nu_3}{k_3}},\label{T0a}\\ T_{\widehat{1}\widehat{2}} &=&
\dot{\phi}^2\,{\frac {\nu_1\nu_2}{k_1k_2}}, \qquad
T_{\widehat{1}\widehat{3}} = \dot{\phi}^2\,{\frac {\nu_1\nu_3}
{k_1k_3}},\qquad T_{\widehat{2}\widehat{3}} = \dot{\phi}^2 \,{\frac
{\nu_2\nu_3}{k_2k_3}}.\label{Tab}
\end{eqnarray}


\end{document}